\documentclass[journal]{IEEEtran}
\usepackage{amssymb,array,algorithm,algorithmic,amsmath,bm,epsfig,epstopdf,graphicx,hyperref,soul,color,subcaption,stfloats,cite,xcolor,multicol,flushend,lipsum,cuted,upgreek,mathtools}
\usepackage{svg}
\usepackage[framemethod=tikz]{mdframed}
\graphicspath{{../Figures/}}

\DeclareMathOperator*{\sinc}{sinc}
\DeclareMathOperator*{\diric}{Diric}

\DeclareMathOperator*{\sfft}{\mathcal{S}}

\DeclareMathOperator*{\otherwise}{otherwise}

\newcommand{\diag}[1]{\ensuremath{\operatorname{diag}}\left\{#1\right\}}

\definecolor{cream}{RGB}{45,  45, 45}
\definecolor{newWhite}{RGB}{162, 183, 185}
\definecolor{cream3}{RGB}{60, 60, 66}
\definecolor{cream2}{RGB}{235,235,235}

\begin{document}
\markboth{IEEE Transactions on Wireless Communications, VOL. XX, NO. XX, \today}{}
\title{Channel Estimation, Interpolation and Extrapolation in Doubly-dispersive Channels}
\author{
\IEEEauthorblockN{Zijun Gong,\,\emph{Member, IEEE},\,Fan Jiang, \emph{Member, IEEE}, Yuhui Song, \emph{Student Member, IEEE}\\
	 Cheng Li,\,\emph{Senior Member, IEEE}, Xiaofeng Tao,\,\emph{Senior Member, IEEE}}
\thanks{
	
Z. Gong is with the IOT Thrust, HKUST (Guangzhou), Guangzhou, Guangdong 511453, China; and the Department of ECE, HKUST, Hong Kong SAR, China (E-mail: gongzijun@ust.hk)}

\thanks{F. Jiang and X. Tao are with the Department of Broadband Communications, Pengcheng Laboratory, Shenzhen 518055, China. X. Tao is also with National Engineering Research Center of Mobile Network Technologies, Beijing University of Posts and Telecommunications, Beijing 100876, China (Email: jiangf02@pcl.ac.cn, taoxf@bupt.edu.cn).}

\thanks{Y. Song is with the Faculty of Engineering and Applied Science, Memorial University of Newfoundland, St. John's, NL, A1B 3X5, Canada (E-mail: yuhuis@mun.ca).}

\thanks{Prof. C. Li is with the School of Engineering Science, Simon Fraser University, Burnaby, BC, V5A 1S6, Canada (E-mail: cheng\_li\_5@sfu.ca)}

}%
\maketitle

\begin{abstract}
The OTFS\,(Orthogonal Time Frequency Space) is widely acknowledged for its ability to combat  Doppler spread in time-varying channels. In this paper, another advantage of OTFS over OFDM\,(Orthogonal Frequency Division Multiplexing) will be demonstrated: much reduced channel training overhead. 
Specifically, the sparsity of the channel in delay-Doppler\,(D-D) domain implies  strong correlation of channel gains in time-frequency\,(T-F) domain, which can be harnessed to reduce channel training overhead through interpolation. An immediate question is how much training overhead is needed in doubly-dispersive channels? A conventional belief is that the overhead is only dependent on the product of delay and Doppler spreads, but we will show that it's also dependent on the T-F window size. The finite T-F window leads to infinite spreading in D-D domain, and aliasing will be inevitable after sampling in T-F domain. Two direct consequences of the aliasing are increased channel training overhead and interference. Another factor contributing to channel estimation error is the inter-symbol-carrier-interference (ISCI), resulting from the uncertainty principle. Both aliasing and ISCI are considered in channel modelling, a low-complexity algorithm is proposed for channel estimation and interpolation through FFT. A large T-F window is necessary for reduced channel training overhead and aliasing, but increases processing delay. Fortunately, we show that the proposed algorithm can be implemented in a pipeline fashion. Further more, we showed that data-aided channel tracking is possible in D-D domain to further reduce the channel estimation frequency, i.e., channel extrapolation. The impacts of aliasing and ISCI on channel interpolation error are analyzed. The spectral efficiency of OTFS and OFDM will be compared by considering the channel estimation error and ISCI. These discussions will shed light on the design of communications systems over doubly-dispersive channels.
\end{abstract}


\section{Introduction}

\subsection{The Legendary Success of OFDM}
From the ADSL\,(Asymmetrical Digital Subscriber Line) and DAB\,(Digital Audio Broadcasting) in the 1990s to WiFi and LTE (Long Term Evolution) in the 2000s, we have witnessed the legendary success of the orthogonal frequency division multiplexing (OFDM) in wireless communications \cite{Weinstein2009}. In 5G and WiFi 7, we are still using this technique for multiplexing and multiple access, i.e., the orthogonal frequency division multiple access (OFDMA). The huge success of OFDM is built upon a very simple channel model, i.e., a linear time-invariant (LTI) model. It is hard to believe that the various characteristics of \emph{wireless channels}, such as dispersion, reflection, multi-path effect, etc. can be very accurately described by an LTI model. With such a simple model, we can use Fourier transform for signal modulation/demodulation on different sub-carriers, because complex sinusoids are eigen-functions of LTI systems. That is to say, when we transmit a sinusoid at $f$ Hz, the receiver will also receive a sinusoid of the identical frequency, although the amplitude and phase will change for sure. In other words, the Doppler effect is totally ignored in the modeling! In \emph{mobile channels}, different paths can have drastically different Doppler frequency shifts (i.e., Doppler spread), and it is very challenging, if not impossible, to compensate for them individually. What makes it worse, different frequency components (or sub-carriers) have different Doppler shifts for ultra-wideband signals, i.e., \emph{Doppler migration} in frequency domain.

\subsection{Will the legend continue?}
A natural question is how did OFDM achieve the great success while ignoring such a fundamental characteristic of mobile channels? As a matter of fact, the basic idea of OFDM was proposed in 1966 \cite{Chang1966}, but its application to mobile communications was not clear until 1985 \cite{Cimini1985}. As we will see in Section \ref{modeling} on channel modeling, the Doppler effect leads to channel variation in time domain. However, for a short time window, the channel is almost static, i.e., the quasi-static channel model, or block fading channel model \cite{Tse2005,Marzetta10}. When OFDM is combined with the multiple-input-multiple-output (MIMO)\,technique, we need to estimate the channel state information\,(CSI) for coherent data detection and spatial multiplexing/multiple access. For the quasi-static channel model, we need to re-estimate the CSI in each frame, and the amount of resources required for channel estimation is proportional to \emph{the number of transmit antennas and the delay spread} \cite{Song21TCOM,Gong19TWC}. An immediate question is how frequently do we need to estimate the CSI? This question can be answered by computing the coherence time, which is inversely proportional to the Doppler spread \cite{Tse2005}. In 5G NR, the channel estimation can be as frequent as four times per slot (each slot contains 14 OFDM symbols) \cite{Dahlman2020}. For reliable communications, the coherence time should be ten times larger than the delay spread\cite{Dean2017,Dean2020}. From another perspective, the Doppler spread leads to inter-carrier-interference (ICI), and the sub-carrier spacing should be at least 100 times the Doppler spread to suppress the ICI to the level of -30 dB, and the bit-error-rate (BER) performance degradation is inevitable \cite{Wang2006}, i.e., error floors will be observed at medium to high SNR regime. Generally, the Doppler spread grows proportionally with carrier frequency and device speed. Naturally, when the device moves faster, the frame length should be reduced, while the amount of resources required for channel estimation remains unchanged, leading to larger overhead. It will eventually become impossible to acquire the CSI in real-time with an affordable cost. This is a fundamental limit of OFDM, and it is deeply rooted in the LTI channel model. Such a problem has been manifesting itself in various forms, and one of the most famous one is the pilot contamination issue in massive MIMO \cite{Marzetta10}.

\subsection{The Promises of OTFS}

The above-mentioned problems of OFDM come from the fact that LTI models cannot capture the time-variant characteristics of the mobile channels. Then can we solve this problem by using a linear time-variant\,(LTV) channel model instead? The orthogonal time frequency space (OTFS) modulation is one of the possible answers to the above mentioned challenges \cite{Monk2016MIMO,Hadani2017Conference,Hadani2018DD}. Other efforts include the affine frequency division multiplexing (AFDM) \cite{Bemani2023}, the orthogonal chirp division multiplexing\,(OCDM) \cite{Xing2016,Omar2021}, the orthogonal delay-Doppler division multiplexing (ODDM) \cite{Lin2022}. In spite of the different waveforms (i.e., \emph{signals}), all these techniques are based on an LTV channel model in the delay-Doppler\,(D-D) domain (i.e., \emph{systems}).  In this paper, we will take OTFS as an example to unveil the huge potentials of signaling techniques over doubly-dispersive channels, in terms of spectral efficiency. 

By employing the D-D domain channel model, OTFS can overcome the shortcomings of OFDM at a price of slightly increased complexity and processing delay. The fundamental reason is that the mobile channel changes much slower in the D-D domain, i.e., the longer \emph{geometric coherence time} \cite{Ramachandran2020,Mishra2022,Raviteja2019}. Then we can estimate the CSI in a much reduced frequency with less cost. From another perspective, although the complex gains of different paths change over time due to the Doppler effect, the way they change can be described by a small number of parameters. It is then possible to estimate these parameters with reduced cost. 
In the following section, the related work on OTFS will be reviewed.

\section{Related Work}

\subsection{Relation Between OFDM and OTFS}
The term OTFS was first proposed in \cite{Monk2016MIMO}, and the model was developed in \cite{Hadani2017Conference}, \cite{Hadani2017mmWave} and \cite{Hadani2018Journal}. The underlying idea of OTFS is that although the Doppler effect leads to phase shift in each path, but the shifting rates of the phases are almost constant in a relatively long period (i.e., linear phase shift over time caused by the Doppler effect)! As a result, we will have time to estimate the channel parameters even in high-mobility scenarios, as we will see later with examples. Another way to understand this is that the mobile channels are generally sparse in the D-D domain, in the sense that the product of delay spread and Doppler spread is much smaller than 1. Thus, there are only a small number of parameters to estimate in the D-D domain.
As the name suggests, OTFS involves signal modeling/processing in three dimensions, i.e., time, frequency and space. The spatial dimension in OTFS comes from the Doppler effect. That is, when a mobile device moves, the Doppler shift leads to a constant phase shift between adjacent time slots on each path. Geometrically, we can synthesize a virtual antenna array  by jointly processing the received signal at different time slots, similar to the concept of SAR (Synthetic Aperture Radar). 

In spite of the difference on mathematical motivations, OTFS is very closely related to OFDM. In \cite{Xia2022}, Dr. Xia pointed out that the two-stage OTFS is basically the vector OFDM\,(VOFDM) he proposed in 2001 \cite{Xia2001}. The modulated signals are indeed very similar, but the motivation of VOFDM was to avoid the spectral nulls and reduce CP length through precoding. In \cite{Raviteja2019WCL}, the authors showed that the OTFS is equivalent to the asymmetric OFDM\,(A-OFDM) for static multi-path channels, a special case of the single-carrier frequency domain equalization (SC-FDE) systems.

\subsection{Diversity Gain}

One of the major motivations of OTFS is the potential diversity gain \cite{Hadani2017Conference}. More than two decades ago, the authors in \cite{Sayeed1999} already pointed out that the temporal variation of wireless channels induced by the Doppler spread can actually be exploited to harness diversity gain in spread-spectrum communications, and the maximum diversity gain is proportional to the product of delay spread and Doppler spread. In \cite{Ma2003}, the authors took one more step and showed that this conclusion holds for not only spread-spectrum systems but also other wideband communications systems in general.

In the pioneer paper on OTFS \cite{Monk2016MIMO}, the authors showed that full diversity gain can be obtained with OTFS modulation/demodulation. The Doppler shift is dependent on angle of arrival, and different arriving paths have different Doppler shifts. With OTFS, we can isolate different paths in the space domain (or Doppler domain), and align their phases. By doing so, we can stop the multi-path components from being destructively combined, and thus eliminate the fading in T-F domain. A diversity gain can thus be obtained. In \cite{Ma2023}, experiments were conducted for high-speed railway systems at 371\,km/h, and a diversity gain of OTFS was verified for a carrier frequency of 450\,MHz. The OTFS can be combined with MIMO to  further enjoy the spatial diversity gain \cite{Surabhi2019TWC}, and proper space-time coding is necessary \cite{Augustine2019}.

\subsection{Channel Estimation for OTFS}
Similar to OFDM, channel estimation is indispensable in OTFS systems, for equalization, multiplexing and multiple access. In \cite{Mohammed2022OTFS}, the authors advocate the D-D domain channel modeling by emphasizing the predictability of the wireless channels in this domain. With OTFS modulation, the authors showed that the equivalent baseband channel in D-D domain is predicable and non-fading, given  that the \emph{crystallization condition} holds. The authors investigated model-based and model-free channel estimation methods \cite{Mohammed2023OTFS}. For the model-based case, they basically assume that there is a small number of distinguishable paths in the D-D domain, and estimate the parameters of each path. For the model-free case, a continuous D-D profile is considered. The model-based scenario is widely considered, and typical algorithms include OMP\,(Orthogonal Matching Pursuit) \cite{Shen2019TSP,Shen2019ICC,Shi2021} and Bayesian learning with expectation-maximization \cite{Liu2020}.

The pilot can be transmitted in different ways. In \cite{Monk2016MIMO}, channel estimation of MU-MIMO (Multi-User-MIMO) systems is considered, and each transmit antenna will sequentially transmit an impulse in the D-D domain. The spacing should be larger than the delay spread and Doppler spread, so that inter-antenna interference can be avoided. In such a case, the required resources for channel estimation is proportional to the \emph{product of transmit antenna number (or user number), delay spread and Doppler spread}. The data and pilot can be isolated in time domain \cite{Ramachandran2018}, or D-D domain \cite{Raviteja2019}. It is even possible to superimpose the pilot on the data in T-F domain \cite{Mishra2022}, and an MMSE\,(Minimum Mean Square Error) channel estimator is developed for single-input-single-output (SISO) communications systems.

\subsection{Contributions and Notations}

In existing work, most of the papers are trying to justify the superiority of OTFS over OFDM by showing that OTFS harnesses the diversity gain in D-D domain, like Dr. Hadani did in his pioneer papers on OTFS. However, we will demonstrate the necessity (not just superiority) of OTFS (or more generally speaking, signaling techniques designed for doubly-dispersive channels) in highly dynamic channels by showing that OFDM is consuming a significant amount of resources on channel estimation, i.e., the spectral efficiency perspective. There are already papers comparing OTFS and OFDM on spectral efficiency. For example in \cite{Gaudio2021} and \cite{Gaudio2022}, the authors compared the achievable rates of OTFS and OFDM in LTV channels, and argued that OTFS has higher efficiency due to shorter cyclic prefix\,(CP). In \cite{Mohammed2021a} and \cite{Mohammed2021b} the spectral efficiency was derived by considering the ISI and ICI, under the assumption of perfect CSI. In this paper, we will consider channel estimation error during the comparison, and show that OTFS has much improved spectral efficiency due to the reduced channel training overhead. The major contributions are summarized below.

\begin{itemize}
	\item We derived the discrete baseband channel model from the continuous channel response. In this process, the implicit approximations/assumptions of the D-D domain channel model will be unveiled, one of which is that the product of bandwidth and frame length must be upper bounded.
	\item The minimum amount of resources required for channel estimation is discussed in the context of general Weyl-Heisenberg systems, and a low-complex channel estimation algorithm based on the fast Fourier transform\,(FFT) is presented to recover the 2D channel response in T-F domain with a small number of training symbols. A pipelined algorithm is proposed to reduce the processing delay, and data-aided channel extrapolation is explored to further reduce channel training overhead.
	\item Apart from noise, two other sources of channel estimation error are unveiled: aliasing in the DD domain resulting from confined time and bandwidth, and ISCI (Inter-symbol-carrier-Interference) due to channel dispersion. These factors are generally ignored in existing work by assuming the bi-orthogonality (no ISCI) and finite D-D spreading (no aliasing). By increasing the time and bandwidth, we can suppress the impact of aliasing, but suffer more from the ISCI.
	\item Comprehensive theoretical and numerical results are presented to compare the spectral efficiencies of OTFS and OFDM. Different from existing work, channel estimation overhead and error are considered in the performance evaluation. These discussions shed light on the design of signaling techniques in doubly-dispersive channels.
\end{itemize}

In the next section, channel modeling will be presented, and channel responses in T-F and D-D domains will be connected. Following that, the channel will be discretized in Section \ref{discretization}, laying the foundation for channel estimation and reconstruction in Section \ref{estimation}. Simulations are presented in Section \ref{numerics}, while Section \ref{conclusions} concludes the paper.

\indent\emph{Notations:} throughout the paper,  $\mathbf{A}[n,m]$ denotes the element of matrix $\mathbf{A}$ on the $n$-th row and $m$-th column. $\mathbf{A}^T$, $\mathbf{A}^H$, and $\mathbf{A}^*$ indicate the transpose, Hermitian transpose, and element-wise conjugate. $\mathcal{I}_N$ is the set of all natural numbers smaller than $N$, i.e., $\mathcal{I}_N=\{0,1,\cdots,N-1\}$. $\oslash$ and $\odot$ indicate element-wise division and multiplication, respectively.

\section{Modeling of Mobile Channels}\label{modeling}
To understand why OTFS outperforms OFDM in highly dynamic channels, we need to review the mobile channels. Consider a carrier frequency of $f_c$, and the general LTV channel model is
\begin{equation}
h(t,\tau)=\sum_{l=0}^{L-1} \beta_{l}e^{-j2\pi f_c\tau_l(t)} \delta(\tau-\tau_l(t)),\label{eqn022}
\end{equation}
where $L$ paths exist between transmitter and receiver. $\beta_l$ is the complex gain of the $l$-th path, and it is almost constant.  $\tau_l(t)$ is the propagation delay of the $l$-th path at time $t$.

Suppose $s(t)$ is the transmitted signal, and the received signal $r(t)$ will be
\begin{equation}
r(t)= \int h(t, \tau)s(t-\tau)d\tau.\label{eqn017}
\end{equation}
To understand the above equation, note that the received signal at time $t$ is the sum of delayed copies of the transmitted signal from every time instant before $t$. The signal transmitted $\tau$ seconds before the current time (i.e., $t$) still has an impact on the current received signal due to channel spreading, quantified by $h(t,\tau)$. Before introducing the channel model in D-D domain, we will first briefly review the root cause of channel variation over time, i.e., the Doppler effect.

\subsection{The Doppler Effect}
Consider a wireless channel with only one path between the transmitter and receiver, and the real-time propagation delay at time $t$ is $\tau_0+at$. Here $a$ describes how fast the propagation delay changes over time, and it will be referred to as the \emph{Doppler scaling factor} in later discussions. Suppose the transmitted signal is $s(t)$, and the channel response is $h(t,\tau)=\beta_0e^{-j2\pi f_c(\tau_0+at)}\delta(\tau - \tau_0 - at)$. The received signal will be
\begin{equation}
		r(t)=\beta_0e^{-j2\pi f_c(\tau_0+at)}s\left((1-a)t-\tau_0\right).
\end{equation}
From this equation, we can see that the signal is first delayed by $\tau_0$ and then scaled in time by a factor of $1-a$. For $a>0$, it means that the receiver is moving away from the transmitter, and the propagation delay is increasing.  On another note, because the received signal is scaled by a factor of $1-a$ in time domain, we will see a scaling factor of $1/(1-a)$ in frequency domain. Suppose the transmitted signal has a frequency of $f$ Hz, the received signal's frequency will be $(1-a)f$, and the Doppler shift is $-af$. 

For acoustic signals, $a$ can be as large as $10^{-2}$ for a speed of several meters per second, while for radio signals in cellular systems, it's at the level of $10^{-8}$ for pedestrians, $10^{-7}$ for vehicles, and upto $10^{-6}$ for high-speed trains. For LEO satellites, $a$ is at the level of $10^{-5}$.  Compared with the OFDM, one fundamental difference of OTFS is its effort to consider the Doppler effect in modeling and exploit it for diversity gain. In the following sub-section, we will derive the baseband equivalent channel model for LTV channels.

\subsection{Approximate Baseband Equivalent Channel Model}
Consider the channel model in \eqref{eqn022}, and the received signal in \eqref{eqn017} can be explicitly written as
\begin{equation}
r(t)=\sum\nolimits_{l}\beta_{l} e^{-j2\pi f_c \tau_{l}(t)}s(t-\tau_{l}(t)).\label{eqn001}
\end{equation}
For a short period of time, we have $\tau_{l}(t)\approx \tau_{l}$, leading to an LTI model. OFDM works very well in such cases.

With OFDM, we can very elegantly remove ISI through Fourier transform at low complexity, but there is a price for that. Note that $\tau_l(t)$'s are assumed to be constant in one data frame, which means the frame cannot be very long. Taking the LTE as an example, one sub-frame lasts for only 1\,ms. In 5G NR (Release 17), a \emph{slot} lasts for 15.625 $\mu$s to 1 ms, with a sub-carrier spacing of 960\,kHz to 15\,kHz, respectively. During each frame/slot, we must use a certain amount of resources for channel estimation, which is part of the overhead we need to pay for doubly-dispersive channels. For high-dynamic scenarios, we need to reduce the frame length, but the amount of resources required for channel estimation in each frame is fixed. As a result, the spectral efficiency will decrease. Intuitively, we need to estimate channel more frequently in wireless channels with higher dynamics, because the channel varies faster. To a certain point, the channel response varies so fast, and it will change before we get a chance to estimate it. Reliable communications will thus become impossible. In extreme cases, the channel is not time-invariant in even one OFDM symbol, and the channel variation will manifest itself through ICI and error floor \cite{Wang2006}.

To solve this problem, note that the LTI channel model is obtained by taking the 0-th order Taylor expansion of the real-time propagation delays, and it cannot capture the time-variant characteristics of the mobile channels. What if we take the first-order Taylor expansion? By doing so, the modeling error will accumulate slower over time and the approximate channel model will stay accurate for a much longer period of time, i.e., the geometric coherence time \cite{Ramachandran2020,Mishra2022}. Thus, channel estimation frequency can be reduced accordingly. Motivated by this idea, we take the propagation delay's first-order Taylor approximation as
\begin{equation}
\tau_l(t)\approx \tau_l + a_l t,\label{eqn010}
\end{equation}
where $a_l$ is the Doppler scaling factor of the $l$-th path.

Consider the approximation in \eqref{eqn010}, and the received signal in \eqref{eqn001} can be approximated by
\begin{equation}
r(t)=\sum_{l}\alpha_l e^{j2\pi\nu_l(t-\tau_{l})}s(t-\tau_{l}-a_lt),
\label{eqn002}
\end{equation}
where $\alpha_l = \beta_le^{j(\phi_l-2\pi f_c \tau_{l}(1+a_l))}$ and 
$\nu_l=-a_lf_c$ are the complex gain and Doppler shift of the $l$-th path, respectively. In rich scattering environment, the D-D domain channel gains  are uncorrelated  Gaussian random variables \cite{Bello1963,Ma2003}.

\begin{figure}[htp]
	\centering
	\includegraphics[width=0.4\textwidth]{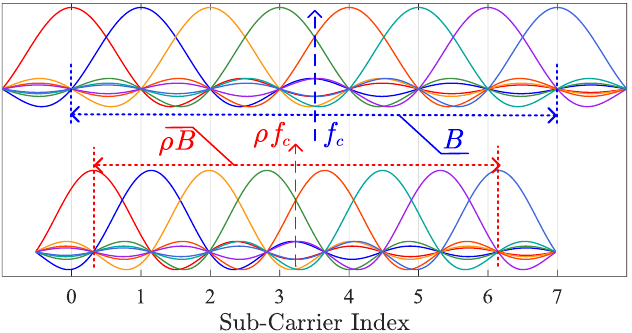}
	\caption{The full Doppler effect, i.e., \emph{shifting} and \emph{scaling}.}
	\label{DoppEffect}
\end{figure}

As we can see, the Doppler effect plays two roles here: first, it shifts the spectrum by $\nu_l$ on the $l$-th path; second, it dilates or compresses the baseband signal by a factor of $1-a_l$ on the $l$-th path.  These two effects are illustrated in Fig. \ref{DoppEffect}, with OFDM modulation over eight sub-carriers. 
\vspace{-0.15in}
The carrier frequency is $f_c$ Hz and the total bandwidth is $B$ Hz. Suppose the scaling factor at the receiver side is $\rho$. The carrier frequency is shifted to $\rho f_c$ (\emph{shifting}), and the total bandwidth is scaled to $\rho B$ (\emph{scaling}). In OFDM, both effects will lead to ICI, but for $B\ll f_c$ the \emph{shifting} effect is much stronger than the \emph{scaling} effect\footnote{Note that $B\ll f_c$ alone does not justify the neglect of the scaling effect.}.

With the first-order expansion, the baseband channel is
\begin{equation}
h(t,\tau) = \sum\nolimits_{l}\alpha_l e^{j2\pi \nu_l(t-\tau_l)}\delta(\tau-\tau_l-a_lt).\label{eqn007}
\end{equation}
Until now we have been talking about the channel model in the  time-delay domain. However, the D-D domain channel model is almost exclusively employed in OTFS-related work. It seems that the above model is very much like the D-D domain channel model by considering propagation delay and Doppler shift (i.e., propagation delay changes over time linearly). However, we will see in the following sub-section that they are fundamentally different.

\subsection{D-D Domain Channel Model}
In \eqref{eqn002}, given that $|a_lt|\ll 1/B$, i.e., the extra delay of the baseband signal caused by the Doppler scaling factor is much smaller than the symbol duration, and we can safely ignore the \emph{scaling} effect on baseband signals. In this case, received signal can be further approximated as
\begin{equation}
r(t)= \sum\nolimits_{l}\alpha_l e^{j2\pi \nu_l(t-\tau_l)}s(t-\tau_l).
\label{eqn003}
\end{equation}
Consider a data frame length of $S$ second, $|a_l|S\ll 1/B$ is equivalent to $BS\ll 1/|a_l|$ for all $l$\cite{Gong2023JSAC}. For wireless channels with radio signals, $1/|a_l|$ is at the level of $10^{6}$ even for high-speed trains. That is to say, we only need to conduct channel estimation once for a resource block of size $BS\ll 3\times 10^{6}$. As comparison, channel estimation is conducted in LTE or 5G NR for a block size of one or two hundred.

If we define the D-D domain channel response as
\begin{equation}
	h(\tau,\nu)=\sum\nolimits_{l}\alpha_l\delta(\nu-\nu_l)\delta(\tau - \tau_l),
\end{equation}
equation \eqref{eqn003} can be rewritten as
\begin{equation}
	\begin{split}
		r(t)=\iint s(t-\tau) h(\tau,\nu)e^{j2\pi \nu(t-\tau)}d\nu
d\tau.\\
	\end{split}\label{eqn004}
\end{equation}
 Equation \eqref{eqn004} is what we commonly see in literature on OTFS. From the above discussions, we can see that the D-D domain channel model is actually the approximation of the first-order Taylor expansion of the mobile channels, by ignoring the scaling effect on baseband signals. Therefore, the D-D domain model is more accurate then the LTI channel model (0-th order), but less accurate than the first-order one in \eqref{eqn007}.

In doubly-dispersive channels, the delay spread and Doppler spread determine how fast channel varies in frequency and time domains, respectively. Suppose Doppler spread is $\nu_d$ (in Hz), and delay spread is $\tau_d$ (in second). For $\tau_d\nu_d<1$, the channel is said to be \emph{underspread}; otherwise, it's \emph{overspread} \cite{Liu2004,Ma2003}. In this paper, we will focus on the underspread channels, which include almost all mobile channels with ratio signals. However, underwater acoustic channels are generally overspread, and the D-D domain channel model is not accurate enough. More specifically, the scaling effect in Fig. \ref{DoppEffect} cannot be ignored anymore, and we have to consider the full Doppler effect. This is why underwater acoustic communications is deemed to be much more challenging than terrestrial communications with radio signals \cite{Liu2004}. 

\begin{figure}[htp]
	\centering
	\includegraphics[width=0.4\textwidth]{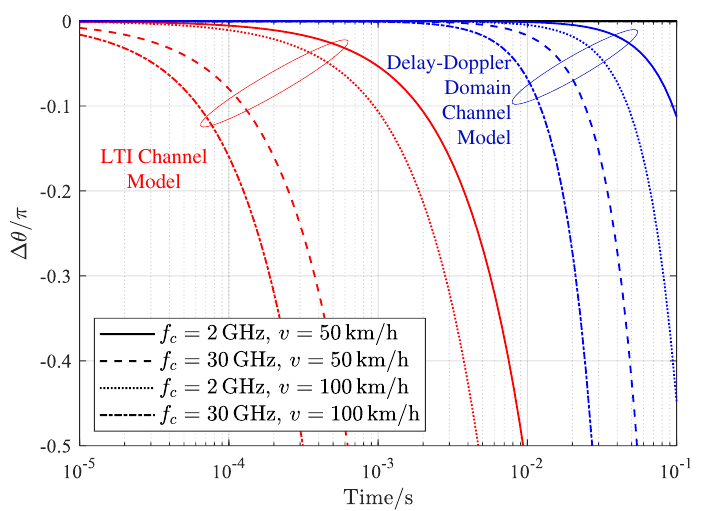}
	\caption{Modelling errors over time.}
	\label{modelErr}
\end{figure}

In Fig. \ref{modelErr}, the channel variations over time are presented for different carrier frequencies and device speeds. The x-axis is time, while the y-axis is the phase change of the complex channel gain. As we have explained, the LTI channel model used in OFDM is the 0-th order Taylor approximation of the accurate one, while the D-D domain channel model is very close to the first-order approximation. Therefore, we can expect the D-D domain channel model to be more accurate. Or equivalently, we can expect this model to stay accurate for a longer period of time. 

As we can see in Fig. \ref{modelErr}, the increase of carrier frequency or user mobility will lead to shorter coherence time for both models. For the same carrier frequency and user mobility, the coherence time of the D-D domain channel model will be longer than that of the LTI model, by two orders of magnitude. Taking the LTE at 2\,GHz as an example, if the vehicle is moving at 50\,km/h, the phase will change 0.05$\pi$ in 1\,ms for the LTI channel model. But for the D-D domain channel model, it takes several tens of milliseconds for the same phase error to happen. This will allow us to comfortably estimate CSI even when the wireless channels are highly dynamic. For example, consider vehicles moving at 100\,m/s, $a_l$ is at the level of $\frac{1}{3}\times 10^{-6}$, which allows a frame length $S\ll \frac{20}{3}$\,s for a bandwidth of 20\,MHz. That is, we only need to estimate the channel every sub-second interval, while in OFDM-based LTE, we need to do that for every 1\,ms! This shows the great potential of OTFS, or signaling techniques engineered for doubly dispersive channels in general, in high-mobility applications.

\section{Channel Discretization}
\label{discretization}

In the previous section, we showed how the D-D domain channel model is related to that in the conventional time-delay domain, and various assumptions in such models are explicitly pointed out. In this section, we will investigate the discrete channel model, laying the foundation for channel interpolation and extrapolation in the next section.

\subsection{Modulation and Demodulation}
Consider a 2D sequence to be transmitted, $\mathbf{X}\in\mathbb{C}^{N\times M}$, with $N$ and $M$ denoting the numbers of symbols in time and frequency domains, respectively. Suppose the symbols are modulated on pulse $g_t(t)$, and the transmitted  signal will be
\begin{equation}
	s(t)=\sum\nolimits_{n,m}\mathbf{X}[n,m]g_t(t-nT)e^{j2\pi mFt},
	\label{eqn005}
\end{equation}
with $T$ and $F$ being the symbol duration and sub-carrier spacing, respectively, satisfying $TF=1$\footnote{In general we have $TF>1$, and example include CP-OFDM and OFDM with guard intervals. The modeling and algorithm design would be similar.}. The total bandwidth is $B=MF$ and the frame length is $S=NT$. In general, we have to make sure $\tau_D< T$ and $\nu_D<F$, which is referred to as the \emph{crystallization condition} in \cite{Mohammed2022OTFS, Mohammed2023OTFS}, and can be easily satisfied for $\tau_D\nu_D\ll 1$.

Then the received signal is 
\begin{equation}
	\begin{split}
		r(t)=&\iint s(t-\tau) h(\tau,\nu)e^{j2\pi \nu(t-\tau)}d\nu
		d\tau\\
		=&\sum\nolimits_{n,m}\mathbf{X}[n,m]\iint h(\tau,\nu)g_t(t-\tau-nT)\cdot\\
		\ &e^{j2\pi(mF+\nu) (t-\tau)}d\tau d\nu.
	\end{split}	\label{eqn009}
\end{equation}
With a matched filter $g_r(t)$ at the receiver side, the received signal is converted to T-F domain as 
\begin{equation}
	y(t,f)=\int_{\tau'} r(\tau')g_{r}(\tau'-t)e^{-j2\pi f(\tau'-t)}d\tau'.
\end{equation}
This operation is referred to as the Wigner transform, and it is basically the short-time Fourier transform with a window of $g_r(t)$ and a phase shift of $2\pi ft$ \cite{Grochenig2000}. Through tedious but straightforward derivations, we can reorganize the output as
\begin{displaymath}
	\begin{split}  
		y(t,f)
		= &  \sum\nolimits_{n,m}\mathbf{X}[n,m]\iint_{\mathcal{D}} h(\tau,\nu)e^{j2\pi(\nu+mF) (t-\tau)}\cdot\\
		\ &A_{g_t,g_r}(t-\tau-nT,f-\nu-mF) d\nu d\tau,
	\end{split}
\end{displaymath} 
where $A_{g_t,g_r}(\tau,\nu)$ is the cross-ambiguity function between $g_t(t)$ and $g_r(t)$, given as
\begin{equation}
	A_{g_t,g_r}(\tau, \nu) = \int g_t(t)g_r(t-\tau)e^{-j2\pi \nu(t-\tau)}dt.
\end{equation} 

\subsection{Discrete Channel Model}
The next step is to discretize the signal in time and frequency domains. Specifically, we sample the 2D signal with an interval of $T$ and $F$ in two dimensions, respectively. The sampled sequence is
\begin{equation}
\begin{split}
\ & \mathbf{Y}[n,m]=y(nT,mF)\\
=&\sum_{\underline{n},\underline{m}}\mathbf{X}[\underline{n},\underline{m}]\iint_{\mathcal{D}}\kappa_{\delta_n,\delta_m(\tau,\nu)}e^{j2\pi (nT\nu-mF\tau)}d\nu d\tau,	
\end{split}\label{eqn018}
\end{equation}
with $\delta_n = n-\underline{n}$ and $\delta_m = m-\underline{m}$. $\kappa_{\delta_n,\delta_m}(\tau,\nu)$ is defined as
\begin{equation}
	\begin{split}
	\ & \kappa_{\delta_n,\delta_m}(\tau,\nu)\\
	=& h(\tau,\nu)e^{-j2\pi(\nu-\delta_m F)\tau
		}A_{g_t,g_r}(\delta_n T-\tau,\delta_m F-\nu),\label{eqn006}
	\end{split}
\end{equation}
which denotes the impact of a transmitted symbol\footnote{To be specific, it is a replica of the transmitted symbol delayed by $\tau$ and shifted by $\nu$ in frequency. The symbol is experiencing spreading due to the multipath effect and Doppler effect. Every replica will have an impact on the received signal, quantified by $h(\tau,\nu)$.} on its neighbours, with a 2D distance of $(\delta_n,\delta_m)$ on the time-frequency grid. From the equations, we can see that the mutual interference among symbols is dependent on both the channel (i.e., \emph{system}) and the transmitting/receiving pulses (i.e., \emph{signal}). 

Apart from interference among symbols, another impact of the delay and Doppler spread is channel variation in T-F domain, and the channel response at the $(n,m)$-th T-F slot is
\begin{displaymath}
	\begin{split}
	\ & \mathbf{H}_{\delta_n,\delta_m}[n,m]=\iint\kappa_{\delta_n,\delta_m}(\tau,\nu)e^{j2\pi (nT\nu-mF\tau)}d\nu d\tau.
	\end{split}
\end{displaymath}
Note that the channel matrix $\mathbf{H}_{\delta_n,\delta_m}$ is sampled from the following 2D channel response:
\begin{displaymath}
	\begin{split}
		H_{\delta_n,
				\delta_m}(t,f)=W(t,f)\iint\kappa_{\delta_n,\delta_m}(\tau,\nu)e^{-j2\pi (f\tau-t\nu)}d\nu d\tau,
	\end{split}
\end{displaymath}
where $W(t,f)$ is a 2D window in time-frequency domain, and the window size is $NT\times MF=B\times S$, i.e.,
\begin{displaymath}
\begin{split}
	W(t,f)
	=\left\{
	\begin{array}{ll}
		1, & t-\frac{T}{2}\in \left(0,NT\right), f-\frac{F}{2}\in (0,MF)\\
		0, &\otherwise.
	\end{array}
	\right.
	\end{split}
\end{displaymath}
Note that $t\in\left(-T/2,(N-1/2)T\right)$ instead of $(0,NT)$, while the range of $f\in\left(-F/2,(M-1/2)F\right)$ instead of $(0,MF)$, so that $W(0,0)$ is well defined.

We then have $\mathbf{H}_{\delta_n,\delta_m}[n,m]=H_{\delta_n,\delta_m}(nT,mF)$.
The T-F window function can be converted to D-D domain through the sympletic Fourier transform (SFT) as
\begin{displaymath}
	\begin{split}
	w(\tau,\nu)=&\iint W(t,f)e^{-j2\pi (t \nu-f\tau )}df dt \\
	= & \frac{\sin(\pi NT\nu)}{\pi \nu}\frac{\sin(\pi MF\tau)}{\pi \tau}e^{-j\pi ((N-1)T\nu-(M-1)F\tau)}.
	\end{split}
\end{displaymath}
This is a 2D sinc function, with a mainlobe width of $\frac{2}{NT}$ (or $\frac{2}{S}$) in $\nu$ and $\frac{2}{MF}$ (or $\frac{2}{B}$) in $\tau$. Then we have
\begin{equation}
	\begin{split}
	H_{\delta_n,\delta_m}(t,f)=& W(t,f)\sfft\{\kappa_{\delta_n,\delta_m}(\tau,\nu)\}\\
	= & \sfft\{w(\tau,\nu)*\kappa_{\delta_n,\delta_m}(\tau,\nu)\},
	\end{split}
\end{equation}
where $*$ denotes convolution, and $\sfft\{\cdot\}$ gives the SFT. 

As a result, the SFT of $H_{\delta_n,\delta_m}(t,f)$ is given as
\begin{equation}
	\begin{split}
	\bar{h}_{\delta_n,\delta_m}(\tau,\nu)=w(\tau,\nu)*\kappa_{\delta_n,\delta_m}(\tau,\nu),\\
	\end{split}
\end{equation}
which is the baseband equivalent channel response in the D-D domain. Note that $\bar{h}_{\delta_n,\delta_m}(\tau,\nu)$ has infinite spreading in the D-D domain, even when $\kappa_{\delta_n,\delta_m}(\tau,\nu)$ is compactly supported.

After 2D sampling in time-frequency domain, i.e., from $H_{\delta_n,\delta_m}(t,f)$ to $\mathbf{H}_{\delta_n,\delta_m}(n,m)$, the channel response in the D-D domain will be periodically extended in both delay and Doppler domains, with a period of $1/F$ and $1/T$, respectively. Let $\tilde{h}_{\delta_n,\delta_m}(\tau,\nu)$ be the periodic extension of $\bar{h}_{\delta_n,\delta_m}(\tau,\nu)$
\begin{equation}
\begin{split}
\tilde{h}_{\delta_n,\delta_m}(\tau,\nu)
=&\sum\nolimits_{k,l} \bar{h}_{\delta_n,\delta_m}(\tau-kT,\nu-lF)\\
= & \kappa_{\delta_n,\delta_m}(\tau,\nu)*\tilde{w}(\tau,\nu).\label{eqn019}
\end{split}
\end{equation}
where $\tilde{w}(\tau,\nu)$ is the periodic extension of $w(\tau,\nu)$ given as
\begin{equation}
	\tilde{w}(\tau,\nu)=\sum\nolimits_{k,l} w(\tau-kT,\nu-lF).
\end{equation}
As has been proved in the appendix, we have
\begin{equation}
\begin{split}
\tilde{w}(\tau,\nu)=&\diric(N,2\pi T\nu)e^{-j\pi (N-1)T\nu}\cdot\\
\ &  \diric(M,2\pi F\tau)e^{j\pi (M-1)F\tau},
\end{split}
\label{eqn016}
\end{equation}
where $\diric(N,\omega)$ is a Dirichlet function (also referred to as the periodic sinc function)\footnote{For $N$ being odd, $\diric(N,\omega)$ gives the $N$-th order Dirichlet kernel.}, defined as
\begin{equation}
\diric(N,\omega)=\frac{\sin N\omega/2}{N\sin \omega/2}.
\end{equation}
$\diric(N,\omega)$ is inherently periodic. Sampling in the T-F domain leads to periodicity in the D-D domain.
We then have
\begin{equation}
\begin{split}
\tilde{h}_{\delta_n,\delta_m}(\tau,\nu)
= & \sum\nolimits_{n,m}\mathbf{H}_{\delta_n,\delta_m}[n,m]e^{-j2\pi( nT\nu-mF\tau)}.
\end{split}
\end{equation}
Direct sampling of $\tilde{h}_{\delta_n,\delta_m}(\tau,\nu)$ leads to a two-dimensional channel matrix in D-D domain: 
\begin{displaymath}
\begin{split}
	\tilde{\mathbf{H}}_{\delta_n, \delta_m}[\tilde{m},\tilde{n}]=& \tilde{h}_{\delta_n, \delta_m}\left(\tilde{m}/B,\tilde{n}/S\right)\\
	=&\sum\nolimits_{n,m}\mathbf{H}_{\delta_n,\delta_m}[n,m]e^{-j2\pi( n\tilde{n}/N-m\tilde{m}/M)}.
	\end{split}
\end{displaymath}
More concisely, we have $\tilde{\mathbf{H}}_{\delta_n,\delta_m} = \sfft_{N,M}\{\mathbf{H}_{\delta_n,\delta_m}\}$,
where $\sfft_{N,M}\{\cdot\}$ denotes the $N\times M$ discrete SFT \footnote{$N\times M$ SFFT is basically $N$-point IDFT vertically and $M$-point DFT horizontally. If the size of target matrix is smaller than $M\times N$, it will be zero-padded to $M\times N$.}. 

The received signal in \eqref{eqn018} can then be rewritten 
\begin{equation}
	\mathbf{Y}[n,m]=\sum\nolimits_{\delta_n, \delta_m}
\mathbf{X}[n-\delta_n,m-\delta_m]\mathbf{H}_{\delta_m,
	\delta_n}[n,m].
\end{equation}
Equivalently, we have
\begin{equation}
	\begin{split}
	\mathbf{Y}=\mathbf{X}\odot \mathbf{H}_{0,0}+\sum_{\delta_n\neq 0, \delta_m\neq 0}\mathbf{X}_{\delta_n, \delta_m}\odot \mathbf{H}_{\delta_n,\delta_m},
	\end{split}
\label{eqn012}
\end{equation}
where $\mathbf{X}_{\delta_n, \delta_m}$ indicates shifted $\mathbf{X}$ by $\delta_n$ and $\delta_m$ in time and frequency domains, respectively.

\subsection{Bi-orthogonality}
\begin{figure}[htp]
	\centering
	\includegraphics[width=0.4\textwidth]{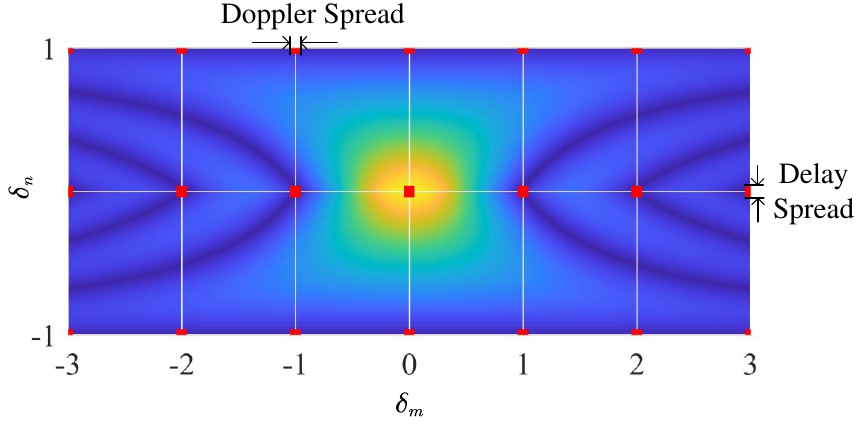}
	\caption{Visualization of bi-orthogonality, ICI and ISI.}
	\label{iscidemo}
\end{figure}

On the right-hand-side of \eqref{eqn012}, there are two parts. The first part is the desired signal, while the second part is the interference in T-F domain, i.e., ISI and ICI, resulting from the imperfect transmitting/receiving pulses and channel dispersion.  Consider OFDM systems with $g_t(t)$ and $g_r(t)$ being identical rectangular pulses, and Fig. \ref{iscidemo} gives the cross-ambiguity function.  
\vspace{-0.15in}

In Fig. \ref{iscidemo}, we can see that the energy of a transmitted symbol will leak to adjacent peers in T-F domain. The red rectangles indicate the delay spread and Doppler spread. With $\tau_D=T/10$ and $\nu_D=F/10$, the product of delay and Doppler spread is equal to 0.01. The cross-ambiguity function will exactly attenuate to zero at integer multiples of $T$ and $F$. For an ideal case with zero delay and Doppler spreads, there will not be ISI and ICI at all, i.e., \emph{bi-orthogonality}. 



The T-F domain channel response is related to these red rectangles. For example, $\mathbf{H}_{0,0}$ can be obtained by zero-padding the red rectangle at $(0,0)$ followed by SFT. All the other rectangles correspond to $\mathbf{H}_{\delta_n,\delta_m}$ for $\delta_n\neq 0$ and $\delta_m\neq 0$. The strength of ISI and ICI is dependent on the area of the red rectangle and also the ambiguity function, i.e., \eqref{eqn006}. This figure gives us an intuitive understanding of how the ISI and ICI are dependent on delay and Doppler spreads. In later discussion, we will refer to the sum of ISI and ICI as ISCI (Inter-Symbol-Carrier-Interference).

$T$ and $F$ should be chosen carefully to minimize the ISCI. In \cite{Liu2004}, the authors investigated optimal pulse design, and they showed that the T-F spread of the modulation pulse should be proportional to the channel spread in D-D domain. For rectangular waveforms in particular, we should choose $T$ and $F$ based on 
\begin{equation}
	T/F=\tau_D/\nu_D.
\end{equation}
We will follow this rule in the simulations.

\section{Channel Interpolation and Extrapolation}
\label{estimation}

Motivated by the fact that the channel changes much slower and thus more predictable in the D-D domain, we will in this section investigate how  we can exploit the predictability for channel interpolation and extrapolation, so that the channel training overhead can be reduced.

 
\subsection{Channel Interpolation with SFT}
By assuming bi-orthogonality (or ignoring the ISCI, equivalently), the signal model can be simplified as
\begin{equation}
	\mathbf{Y}[n,m]=\mathbf{H}[n,m]\mathbf{X}[n,m],\label{eqn008}
\end{equation}
where $\mathbf{H}=\mathbf{H}_{0,0}$  and we ignored the subscript for  conciseness. Similarly we represent $\tilde{\mathbf{H}}_{0,0}$ with $\tilde{\mathbf{H}}$ in D-D domain, $\kappa(\tau,\nu)$ for $\kappa_{0,0}(\tau,\nu)$, and  $\tilde{h}(\tau,\nu)$ for $\tilde{h}_{0,0}(\tau,\nu)$.
This is actually a 2D flat fading channel, and we can see a strong resemblance between this model and the LTI model used in OFDM. 

\begin{figure}[htp]
	\centering
	\includegraphics[width=0.4\textwidth]{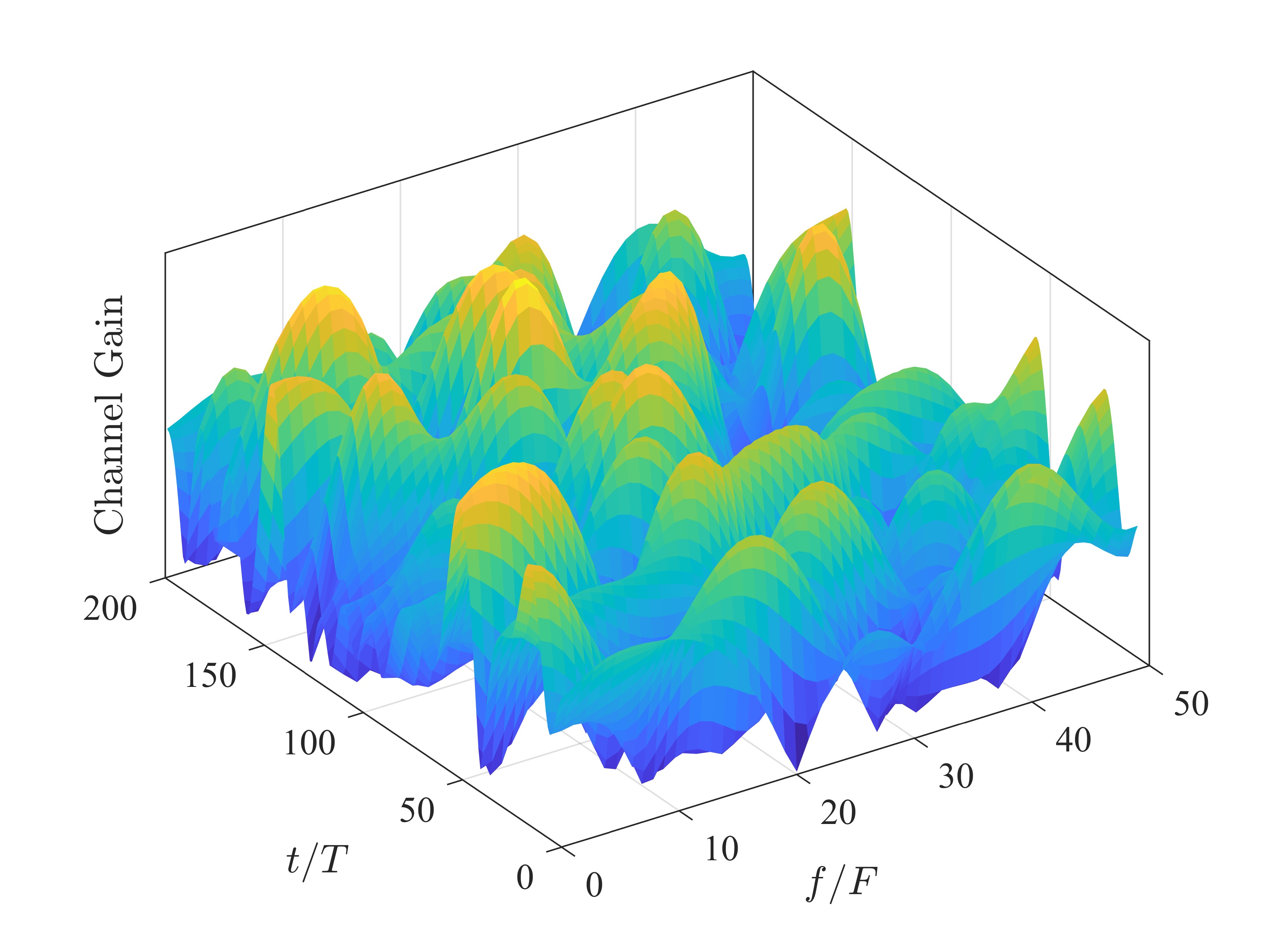}
	\caption{Demonstration of the doubly-selective fading channel in time-frequency domain, with bi-orthogonality.}
	\label{selecFad}
\end{figure}

Consider a vehicular speed at 100 m/s, and the WSSUS channel model, Fig. \ref{selecFad} shows one realization of the wireless channel in T-F domain.
\vspace{-0.2in}
In Fig. \ref{selecFad}, the carrier frequency is 30\,GHz, with a sub-carrier spacing of 200\,kHz, total bandwidth of 10\,MHz, and 1\,ms frame length (or 200 symbols). As we can see, the channel changes very fast over both time and frequency, and  the channel gains can be significantly different even between adjacent T-F slots. This example demonstrates the \emph{necessity} of OTFS in highly dynamic channels.

As we can see, the T-F resource block is divided into small blocks of unit area, and all the sub-channels have different complex gains. But similar to OFDM, these channel gains are correlated, and it is possible to recover the complete channel response with a small number of samples. In other words, $\mathbf{H}[n,m]$ is sparse. Although we cannot see the sparsity of the channel response in T-F domain directly,  note that $\mathbf{H}[n,m]$ is obtained from $\tilde{\mathbf{H}}$ in D-D domain, and the latter is inherently sparse. For a delay spread of $\tau_d$ and a Doppler spread of $\nu_d$, most of elements in $\tilde{\mathbf{H}}$ are close to zero. Specifically in \eqref{eqn019}, $\kappa(\tau,\nu)$ is zero outside an area of $\tau_D\times \nu_D$, while the mainlobe width of $\tilde{w}(\tau,\nu)$ is $\frac{2}{B}\times \frac{2}{S}$. 

\begin{figure}[htp]
	\centering
	\includegraphics[width=0.45\textwidth]{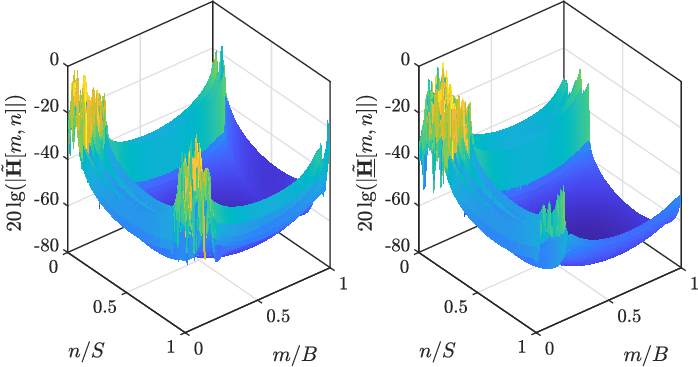}
	\caption{Rotation from $\tilde{\mathbf{H}}$ to $\underline{\tilde{\mathbf{H}}}$.}
	\label{rotation}
\end{figure}

Consider the primary period of $\tilde{h}(\tau,\nu)$, i.e., $\tau\in[0,T)$ and $\nu\in[0,F)$, and $\tilde{h}(\tau,\nu)$ will be close to zero outside the following area
\begin{equation}
	\begin{array}{ll}
	\ & \tau \in \left(0, \tau_D+\frac{1}{B}\right)\cup \left(T-\frac{1}{B}, T\right),\\
	\ & \nu \in (0,\frac{\nu_D}{2}+\frac{1}{S})\cup \left(F-\frac{\nu_D}{2}-\frac{1}{S}, F\right).
	\end{array}
\end{equation}
$\tilde{\mathbf{H}}$ is sampled from $\tilde{h}_{\delta_n,\delta_m}(\tau,\nu)$ with an interval of $1/B$ in $\tau$ and $1/S$ in $\nu$. The amplitude of $\tilde{\mathbf{H}}$ is illustrated in the left-hand-side of Fig. \ref{rotation}.
\vspace{-0.1in}

As we can see from Fig. \ref{rotation}, $\tilde{\mathbf{H}}$ has large values in its four corners. For notational convenience in later discussions, we circularly rotate $\tilde{\mathbf{H}}$ by defining $\underline{\tilde{\mathbf{H}}}$ as 
\begin{equation}
	\underline{\tilde{\mathbf{H}}}[m,n] = \tilde{\mathbf{H}}[(m-1)_M,(n-\breve{N}/2)_N],
\end{equation}
where $\breve{N}\geq \lceil \nu_d S + 2\rceil$ and $(a)_b=\mod(a,b)$. Then it follows that 
\begin{equation}
	\underline{\tilde{\mathbf{H}}}[m,n]\approx 0 \text{ for } m<\breve{M}, n<\breve{N}.
\end{equation}
with $\breve{M}\geq \lceil \tau_d B + 2\rceil$. In the right-hand-side of Fig. \ref{rotation}, we can see that the non-zero values are now more concentrated in just one corner of the channel matrix.
We thus have the SFT of $\underline{\tilde{\mathbf{H}}}$ as
\begin{equation}
	\underline{\mathbf{H}}=\sfft\nolimits_{M,N}\{\underline{\tilde{\mathbf{H}}}\}=\mathbf{D}_N\mathbf{H}\mathbf{D}_M^*,\label{eqn021}
\end{equation}
with $\mathbf{D}_K=\diag{\left[1, e^{j\omega_K}, e^{j2\omega_K},\cdots, e^{j(K-1)\omega_K}\right]}$.

%

The sparsity of $\underline{\tilde{\mathbf{H}}}$ (or $\tilde{\mathbf{H}}$) suggests that we can down-sample the channel response in T-F domain without much information loss. Therefore, only a small number of samples are required for channel estimation. At the receiver side, we can estimate channel gains on these chosen slots, and estimate  others through interpolation, like we did in OFDM systems.

To start with, we down-sample $\underline{\mathbf{H}}$ by a factor of $L_N$ and $L_M$ in time and frequency domains\footnote{Without loss of generality, we assume that $M/L_M$ and $N/L_N$ are integers. This can be guaranteed by choosing $T$ and $F$ properly.}, respectively.  $L_N\geq N/\breve{N}$ and $L_M\geq M/\breve{M}$ should be guaranteed, and without loss of generality, we assume $L_N= N/\breve{N}$ and $L_M= M/\breve{M}$. The down-sampled version is $\underline{\breve{\mathbf{H}}}$ as
\begin{equation}
	\underline{\breve{\mathbf{H}}}[\breve{n},\breve{m}]=\underline{\mathbf{H}}[\breve{n}L_N , \breve{m}L_M].
\end{equation} 
The down-sampling in T-F domain leads to periodic extension in the D-D domain, given as 
\begin{equation}
		\underline{\tilde{\mathbf{H}}}_p[\breve{m},\breve{n}]=\sum\nolimits_{\delta_m,\delta_n} \underline{\tilde{\mathbf{H}}}[\delta_mL_M +\breve{m}, \delta_nL_N+\breve{n}],
\end{equation}
and we can easily verify 
\begin{equation}
	\begin{split}
		\underline{\breve{\mathbf{H}}}=&\sfft\nolimits_{\breve{M},\breve{N}}\left\{\tilde{\underline{\mathbf{H}}}_p\right\}\\
		\approx & \sum_{\tilde{n}=0}^{\breve{N}-1}\sum_{\tilde{m}=0}^{\breve{M}-1} \underline{\tilde{\mathbf{H}}}[\tilde{m},\tilde{n}]e^{-j(  \breve{m}\tilde{m}\omega_{\breve{M}}-\breve{n}\tilde{n}\omega_{\breve{N}})},\label{eqn023}
	\end{split}
\end{equation} 
with $\omega_{\breve{M}}=2\pi/\breve{M}$ and $\omega_{\breve{N}}=2\pi/\breve{N}$. The approximation comes from the neglect of the sidelobes of in D-D domain. More concisely, we approximately have
\begin{equation}
	\underline{\breve{\mathbf{H}}}\approx\sfft\nolimits_{\breve{M},\breve{N}}\{\underline{\tilde{\mathbf{H}}}[0:\breve{M}-1],0:\breve{N}-1\}.
\end{equation}
We can thus recover $\underline{\tilde{\mathbf{H}}}$ from $\underline{\breve{\mathbf{H}}}$ first, and then reconstruct $\mathbf{H}$ from $\underline{\tilde{\mathbf{H}}}$, given as
\begin{equation}
	\underline{\mathbf{H}}=\sfft\nolimits_{M,N}\{\sfft\nolimits_{\breve{N},\breve{M}}\{\underline{\breve{\mathbf{H}}}\}\}.\label{eqn020}
\end{equation}

In the above discussions, we mentioned seven different channel representations, four in T-F domain and three in D-D domain. For symmetry, we introduce $\tilde{\mathbf{H}}_p$ as the periodic extension of $\tilde{\mathbf{H}}$. The following figure shows how these eight channel representations are related. 
\vspace{-0.1in}
\begin{figure}[htp]
	\centering
	\includegraphics[width=0.3\textwidth]{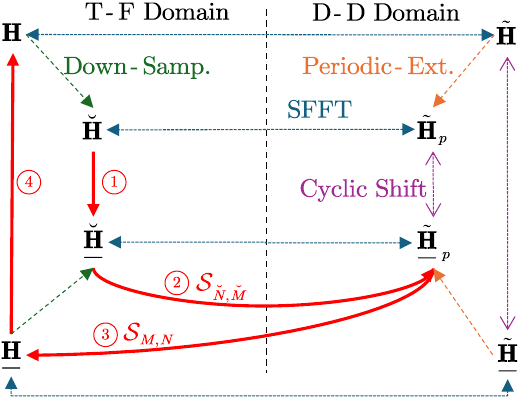}
	\caption{Eight channel representations.}
	\label{flow}
\end{figure}

The red arrows in Fig. \ref{flow} indicate four essential steps in channel interpolation. We start with $\breve{\mathbf{H}}$, and obtain $\underline{\breve{\mathbf{H}}}$ through phase rotation given below
\begin{equation}
	\underline{\breve{\mathbf{H}}}=\breve{\mathbf{D}}_N\breve{\mathbf{H}}\breve{\mathbf{D}}_M^*.
\end{equation}
where $\breve{\mathbf{D}}_N$ and $\breve{\mathbf{D}}_M$ are sampled from $\mathbf{D}_N$ and $\mathbf{D}_M$, respectively. Then  step 2 and 3 are SFT given in \eqref{eqn020}. The last step is phase rotation again through \eqref{eqn021}.

\noindent \emph{Remarks:} Based on the above discussions,  the minimum number of T-F slots required for channel estimation is approximately
\begin{equation}
	\lceil \tau_DB+2\rceil \lceil \nu_D S+2\rceil\approx BS\tau_D\nu_D+4.
\end{equation}
The ratio of channel training overhead is approximately
\begin{equation}
\frac{BS\tau_D\nu_D+4}{NM}=\tau_D\nu_D + \frac{4}{BS}.
\end{equation}
When $BS$ is very large, the overhead ratio is $\tau_D\nu_D$. For terrestrial communications with radio signals, the channel is generally underspread with $\tau_D\nu_D\ll 1$. We can then estimate the CSI with a small training overhead. More resources for channel estimation leads to more accurate CSI, but the increased overhead also means reduced efficiency. However, if $BS$ is not large enough, the training overhead can be significantly larger than $\tau_D\nu_D$. 

\noindent \emph{Remarks:} The above discussions hold for both discrete and continuous D-D profiles. For discrete D-D channel model, these discussions hold even when the Doppler shifts and delays of different paths are not exactly on the D-D grids in general.

\subsection{A Pipelined Implementation}
The previous sub-section shows that channel interpolation is possible in doubly-dispersive channels. However, this means we have to jointly process a large data block to harness the sparsity of channel response in the D-D domain. We can of course reduce $B\times S$, but the reduced window means increased spreading of $\tilde{w}(\tau,\nu)$ and $\tilde{h}(\tau,\nu)$ in D-D domain, leading to larger channel training overhead and also larger truncation error (aliasing) in \eqref{eqn023}. In this sub-section, we will present a pipelined implementation for channel interpolation algorithm, so that the processing delay can be reduced without sacrificing the D-D domain resolution, or inducing too much aliasing.

Based on the pilot symbols transmitted at time $\breve{n}L_NT$ for $\breve{n}\in \mathcal{I}_{\breve{N}}$, we can get the CSI for time $[0,NT)$ through channel interpolation, as we have discussed in the previous sub-section. Then, pilot can be inserted at time $NT=\breve{N}L_NT$, and channel interpolation can then be conducted for time duration $[T, (N+1)T)$ through FFT. 

\begin{figure}[htp]
\centering
\includegraphics[width=0.4\textwidth]{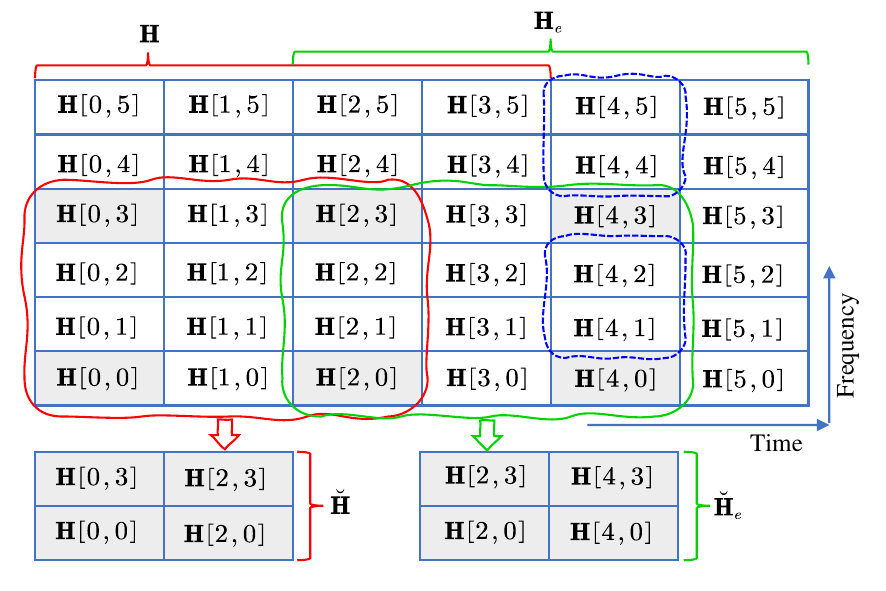}
\caption{Illustration of pipeline implementation of the channel interpolation, with $N=4$, $M=6$, $L_N=2$, $L_M=3$.}	
\label{chanExtrap}
\end{figure}

To understand why this is possible, consider channel interpolation for $t\in[\Delta_nT, (N+\Delta_n) T))$ and $f\in[0, MF)$, with $\Delta_n$ being a natural numbers. Still consider $n\in \mathcal{I}_N$ and $m\in \mathcal{I}_M$, the channel gains on the T-F grid are given by
\begin{equation}
	\mathbf{H}_e[n,m]=\iint\kappa(\tau,\nu)e^{j2\pi ((n+\Delta_n)T\nu-mF\tau)}d\nu d\tau. 
\end{equation}
Again, note that the channel matrix $\mathbf{H}_e$ is sampled from the following 2D channel response:
\begin{displaymath}
		H_e(t,f)
		=W(t-\Delta_nT,f)\iint\kappa(\tau,\nu)e^{j2\pi (t\nu-f\tau)}d\nu d\tau,
\end{displaymath}
with $\mathbf{H}_e[n,m]=H_e((n+\Delta_n)T,mF)$.
Then we have
\begin{equation}
	\begin{split}
	H_e(t,f)=& W(t-\Delta_nT,f)\sfft\{\kappa(\tau,\nu)\}\\
	= & \sfft\{e^{-j2\pi \Delta_nT\nu}w(\tau,\nu)*\kappa(\tau,\nu)\}.
	\end{split}
\end{equation}
As a result, the SFT of $H_e(t,f)$ is given as
\begin{equation}
	\bar{h}_e(\tau,\nu)=e^{-j2\pi \Delta_nT\nu}w(\tau,\nu)*\kappa(\tau,\nu),
\end{equation}
Again, sampling in the T-F domain leads to periodical extension of the D-D domain channel response:
\begin{displaymath}
	\tilde{h}_e(\tau,\nu)=\sum_{k,l}\bar{h}(\tau-kT,\nu-lF)=e^{-j\pi 2\Delta_nT\nu}\tilde{h}(\tau,\nu).
\end{displaymath}
Similar to $\tilde{h}(\tau,\nu)$, $\tilde{h}_e(\tau,\nu)$ is also mostly zero outside a small region of around $\tau_D\nu_D$. We can thus use the same interpolation technique presented in the previous sub-section, and the basic idea is shown in Fig. \ref{chanExtrap}.
\vspace{-0.15in}

In Fig. \ref{chanExtrap}, we first conduct channel interpolation for $n\in\mathcal{I}_3$ and $m\in\mathcal{I}_5$, based on pilots at $n\in\{0,2\}$ and $m\in\{0,3\}$, indicated by the red bracket. Then when we receive pilot at $n=4$, we will be able to combine the pilots from $n=2$ and $n=4$ for channel interpolation from time $n=2$ to $n=5$, indicated by the green bracket. 

\subsection{Channel Extrapolation and Tracking}
\begin{figure}[htp!]
	\centering
	\includegraphics[width=0.4\textwidth]{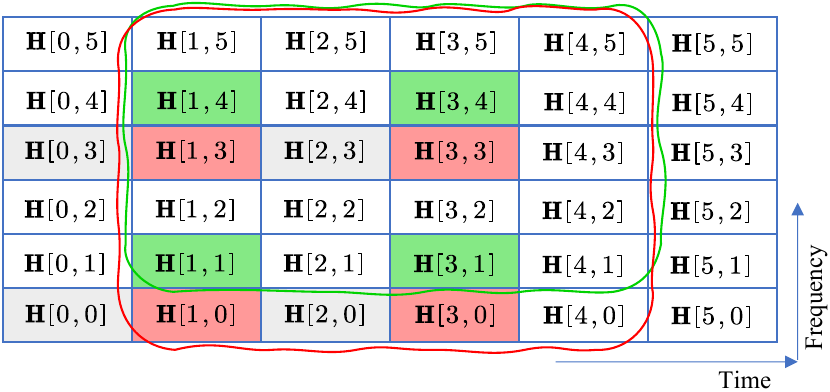}
	\caption{Data-aided channel extrapolation. }	
	\label{dataAidedChanExtrap}
\end{figure}

In Fig. \ref{chanExtrap}, note that the data received at  $n=4$, i.e., encircled by the blue curves, cannot be demodulated immediately, because they have to wait for the pilot at time $4$. However, based on the previous discussions, we should be able to use the previously estimated CSI for channel prediction. 
Specifically, we can employ the estimated CSI from time $1$ and $3$ for channel interpolation for time $1$ to $4$. From a different point of view, this is extrapolation. This also implies the possibility of channel tracking, and we further reduce the channel training overhead by inserting pilot less frequently. The idea is illustrated in Fig. \ref{dataAidedChanExtrap}.
\vspace{-0.15in}

With pilot transmitted at time $n=0$ and $2$, frequency $m=0$ and $3$, channel interpolation can be conducted for time $0\leq n\leq 3$ and frequency $0\leq m\leq 5$. Then, we can use channel gains at $n\in\{1,3\}$, $m\in\{0,3\}$, i.e., the slots indexed by red, for channel interpolation between $n=1$ and $4$, encircled by red. We can thus estimate the channel gains at time 4 and 5, without waiting for the pilot. The processing delay can thus be further reduced to one symbol duration. 

\begin{figure*}
		\begin{equation}
			\begin{split}
				\underline{\hat{\mathbf{H}}}
				=\sfft\nolimits_{M,N}\Big\{\underline{\tilde{\mathbf{H}}}+\underbrace{\left(\mathbf{W}_{\breve{N}\breve{M}}-\mathbf{1}_{NM}\right)\odot\underline{\tilde{\mathbf{H}}}}_{I_0}+\underbrace{\mathbf{W}_{\breve{N}\breve{M}}\odot\sum_{\breve{n}\neq 0,\breve{m}\neq 0}\underline{\tilde{\mathbf{H}}}^{(\breve{n}L_N,\breve{m}L_M)}}_{I_1}\Big\}+\underbrace{\sfft\nolimits_{M,N}\left\{\sfft\nolimits_{\breve{N},\breve{M}}\{\underline{\breve{\mathbf{H}}}_I\}\right\}}_{I_2}\\
			\end{split}
			\label{eqn015}
		\end{equation}
	\hrule
\end{figure*}

A risk of data-aided channel tracking is the error propagation issue, and we can reduce the risk through averaging. For example in Fig. \ref{dataAidedChanExtrap}, the channel gains indicated by green can also help us to reconstruct the channel at time 4. An average can be taken over the predictions obtained from red or green for better reliability. More generally, we use CSI collected from time 
$n\in  \Delta_n+L_N\times \mathcal{I}_{\breve{N}}$ for channel prediction at $n=N$, with $\Delta_n = 1, 2,\cdots,L_N-1$.


\subsection{Error Analysis}

The interpolation inevitably leads to error, because the D-D domain channel has infinitely large spread due to the finite support in T-F domain, manifested by the 2D sinc function in D-D domain. Besides, the bi-orthogonality of the signal no longer holds after going through the doubly-dispersive channel. In this sub-section, we will try to quantify the channel estimation errors resulting from aliasing and ISCI.

Define $\breve{\mathbf{X}}$ and $\breve{\mathbf{Y}}$ as the transmitted and received pilot signals
\begin{displaymath}
		\breve{\mathbf{X}}[\breve{n}, \breve{m}] =  \mathbf{X}[\breve{n}L_N, \breve{m}L_M],\
		\breve{\mathbf{Y}}[\breve{n}, \breve{m}] =  \mathbf{Y}[\breve{n}L_N, \breve{m}L_M].
\end{displaymath}
The received pilot is
\begin{equation}
	\begin{split}
	\breve{\mathbf{Y}}=\sum_{\delta_n, \delta_m}\breve{\mathbf{X}}_{\delta_n, \delta_m}\odot \breve{\mathbf{H}}_{\delta_n,\delta_m},
	\end{split}
\end{equation}
where $\breve{\mathbf{H}}_{\delta_n,\delta_m}$ and $\breve{\mathbf{X}}_{\delta_n, \delta_m}$ are down-sampled versions of $\mathbf{H}_{\delta_n,\delta_m}$ and $\mathbf{X}_{\delta_n,\delta_m}$, respectively, by a factor of $L_N$ in time and $L_M$ in frequency. Then correspondingly, if we try to recover the complete T-F response, aliasing will be inevitable. The estimated channel matrix will be
\begin{equation}
	\begin{split}
		\hat{\breve{\mathbf{H}}}=\breve{\mathbf{Y}}\oslash \breve{\mathbf{X}}=&\sum_{\delta_n, \delta_m}\breve{\mathbf{X}}_{\delta_n, \delta_m}\odot \breve{\mathbf{H}}_{\delta_n,\delta_m}.
	\end{split}
\end{equation}
Then we can rotate the channel matrix as indicated by step 1 in Fig. \ref{flow}
\begin{equation}
	\begin{split}
		\underline{\hat{\breve{\mathbf{H}}}}=&\mathbf{D}_{\breve{M}}\hat{\breve{\mathbf{H}}}\mathbf{D}_{\breve{N}}^*\\
		=&\underline{\breve{\mathbf{H}}}+\underbrace{\sum_{\delta_n\neq 0, \delta_m\neq 0} \mathbf{D}_{\breve{M}}(\breve{\mathbf{H}}_{\delta_n,\delta_m}\odot\breve{\mathbf{X}}_{\delta_n, \delta_m}\oslash\breve{\mathbf{X}})\mathbf{D}_{\breve{N}}^*}_{\underline{\breve{\mathbf{H}}}_{I}}.
	\end{split}
\end{equation}

The SFT of $\underline{\breve{\mathbf{H}}}$ will be is periodic extension of $\underline{\tilde{\mathbf{H}}}$, i.e., $\underline{\tilde{\mathbf{H}}}_{p}$ given by
\begin{equation}
\hat{\underline{\tilde{\mathbf{H}}}}_{p}=\underline{\tilde{\mathbf{H}}}_p+\sfft\nolimits_{\breve{N},\breve{M}}\{\underline{\breve{\mathbf{H}}}_I\}.
\end{equation}
The step 3 in Fig. \ref{flow} give us
\begin{equation}
	\begin{split}
	\hat{\underline{\mathbf{H}}}=\sfft\nolimits_{M,N}\left\{\underline{\hat{\tilde{\mathbf{H}}}}_p\right\}.
	\end{split}
\end{equation}

Through the above steps we recover the complete channel response in T-F domain, and the exact expression of the recovered CSI is given by \eqref{eqn015}. $\mathbf{1}_{NM}$ is a all-one matrix of size $N\times M$, and  $\mathbf{W}_{\breve{N}\breve{M}}\in\mathbb{R}^{N\times M}$ is given as
\begin{equation}
	\mathbf{W}_{\breve{N}\breve{M}}[n,m]=\left\{
	\begin{array}{ll}
		1, & n\leq \breve{N}-1,\ m\leq \breve{M}-1, \\
		0, & \otherwise.\\
	\end{array}
	\right.
\end{equation} 
The first part $\underline{\tilde{\mathbf{H}}}$ is the desired CSI. $I_0$ results from truncation due to the limited T-F resource. $I_1$ is the aliasing resulting from down-sampling of the T-F channel response, and the infinite spreading of the sinc function, resulting from the finite window in T-F domain. The errors in both parts can be reduced by using a larger $B\times S$, or increased pilot resource, i.e., $\breve{N}$ and $\breve{M}$.

The third part $I_2$ is the ISCI. This part grows when we increase $B\times S$. Apparently, the ripples of the 2D sinc will diminish for larger $B\times S$. However, the ISCI will increase. So the $B$ and $S$ should be carefully chosen to achieve balance between these two types of errors.

\section{Numerical Evaluations}\label{numerics}
In this section, we will evaluate the theoretical analyses presented in the previous discussions, based on the WSSUS channel model. Consider a carrier frequency $f_c=30$\,GHz, and a sub-carrier spacing of 200\,kHz, or a symbol duration of 5\,$\upmu$s. QPSK modulation is considered throughout the simulations.



\subsection{ISCI from Delay-Doppler Spreading}
\begin{figure}[htp]
	\centering
	\includegraphics[width=0.4\textwidth]{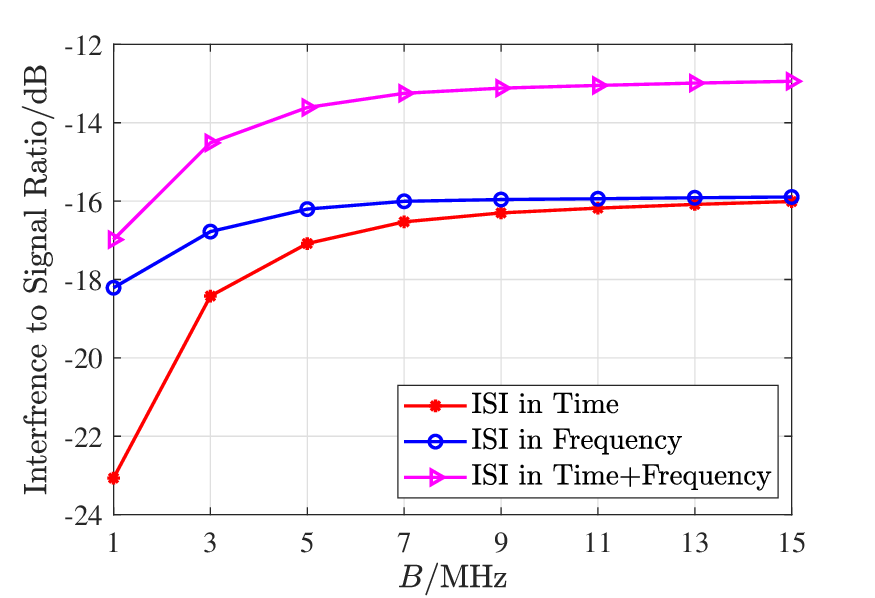}
	\caption{ISCI for different bandwidth.}
	\label{isiB}
\end{figure}

OFDM is based on the LTI channel model, while OTFS is built upon the D-D domain channel model and the bi-orthogonality. Both models suffer from modeling error, leading to system errors in OFDM and OTFS. For OFDM, the Doppler spread leads to ICI, and also time-domain channel variation; for OTFS, the modeling error results from the time-frequency spreading of the cross-ambiguity function of the transmitting/receiving pulses, as we can see from Fig. \ref{iscidemo}. Fig. \ref{isiB} presents the ISI and ICI, for a delay spread of $\tau_D=300/c=1\upmu$s and a Doppler spread of $20$\,kHz.

In Fig. \ref{isiB}, the bandwidth varies from 1 to 15 MHz. As we can see, the ISI and ICI are both increasing as the bandwidth increases, and they will gradually level off. Because most interference is from adjacent sub-carriers. As the bandwidth increase, the remote sub-carriers will have a weaker impact. The ISI and ICI are at the same level. In OFDM, by adding CP, we will be able to remove the ISI, but the ICI will be inevitable. This is the main source of performance degradation in OFDM. In OTFS, we can also remove the ISI by adding CP. However, for high-mobility applications, adding CP will not necessarily boost system performance. We can eliminate ISI by paying the overhead of CP, but the ICI adheres. The ISCI can thus be reduced by 3\,dB, which leads to a spectral efficiency of 1\,bps/Hz in high SNR regimes, which cannot necessarily compensate the overhead of CP.

\begin{figure}[htp]
	\centering
	\includegraphics[width=0.45\textwidth]{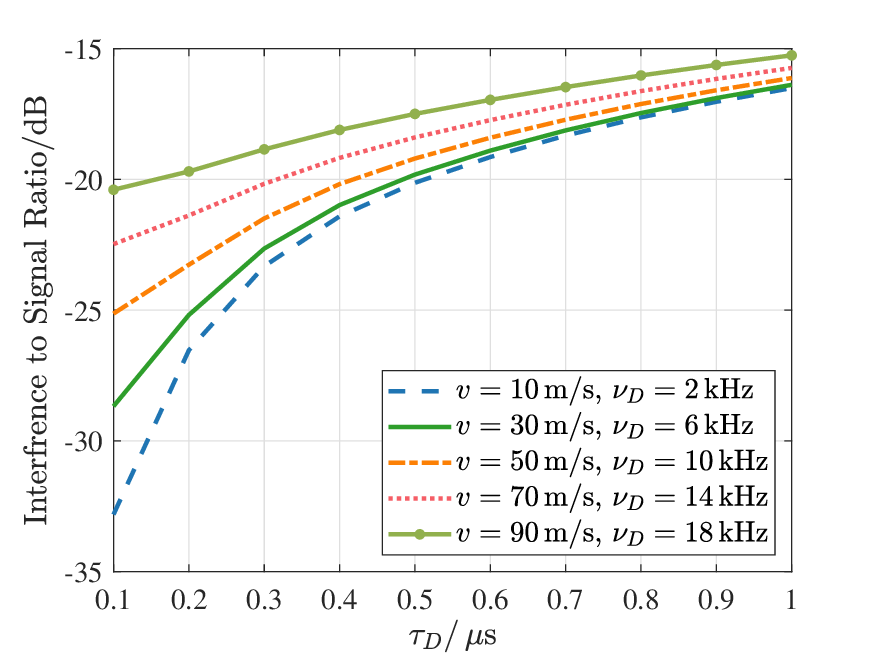}
	\caption{ICI for different D-D spreads.}
	\label{isi_tauDnuD}
\end{figure}

The ISCI is apparently dependent on the delay and Doppler spreads. Larger spreads will lead to increased ISCI. In Fig. \ref{isi_tauDnuD}, the interference to noise ratio (ISR) is presented for different delay spreads and Doppler spreads.

As we can see from Fig. \ref{isi_tauDnuD}, the $\tau_D$ varies from 0.1 to 1\,$\upmu$s, while the Doppler spread varies from 2 kHz to 18 kHz, corresponding vehicular speeds of 10 m/s  and 90 m/s, respectively. The ISCI is at the level of -30 to -15\,dB, which cannot be ignored. For medium- to high-SNR regime, the bi-orthogonality assumption does not hold anymore.

\subsection{Aliasing From T-F Windowing}
\begin{figure}[htp]
	\centering
	\includegraphics[width=0.45\textwidth]{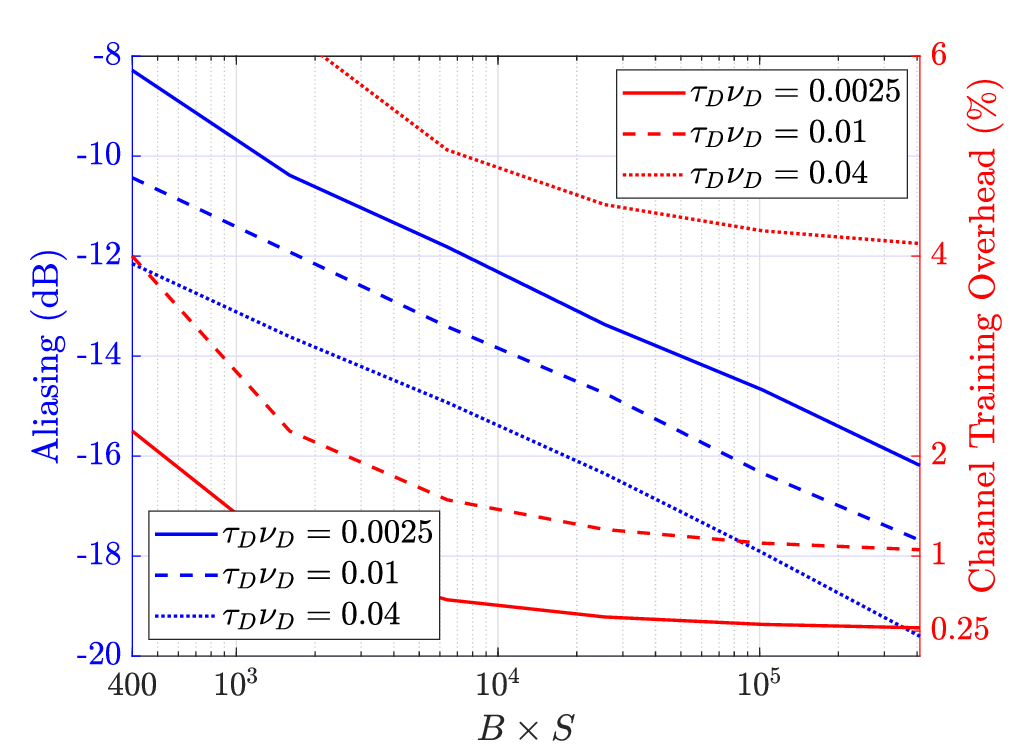}
	\caption{Aliasing introduced by T-F windowing.}
	\label{alias}
\end{figure}

Apart from ISCI, aliasing also contributes to channel interpolation/extrapolation error. The results are presented in n Fig. \ref{alias}.

As we can see, when we increase $B\times S$, the aliasing will decrease, and the channel training overhead will decrease, at a price of higher computational complexity. The processing delay can be controlled through the proposed pipeline implementation. From the numerical results, we can see that $B\times S$ should be at the level of $10^5$. The channel training overhead will gradually converge to $\tau_D\nu_D$. For example, the channel training overhead is approximately four percent for $\tau_D\nu_D=0.04$.

\subsection{Channel Interpolation Error}

\begin{figure}[htp]
	\centering
	\includegraphics[width=0.45\textwidth]{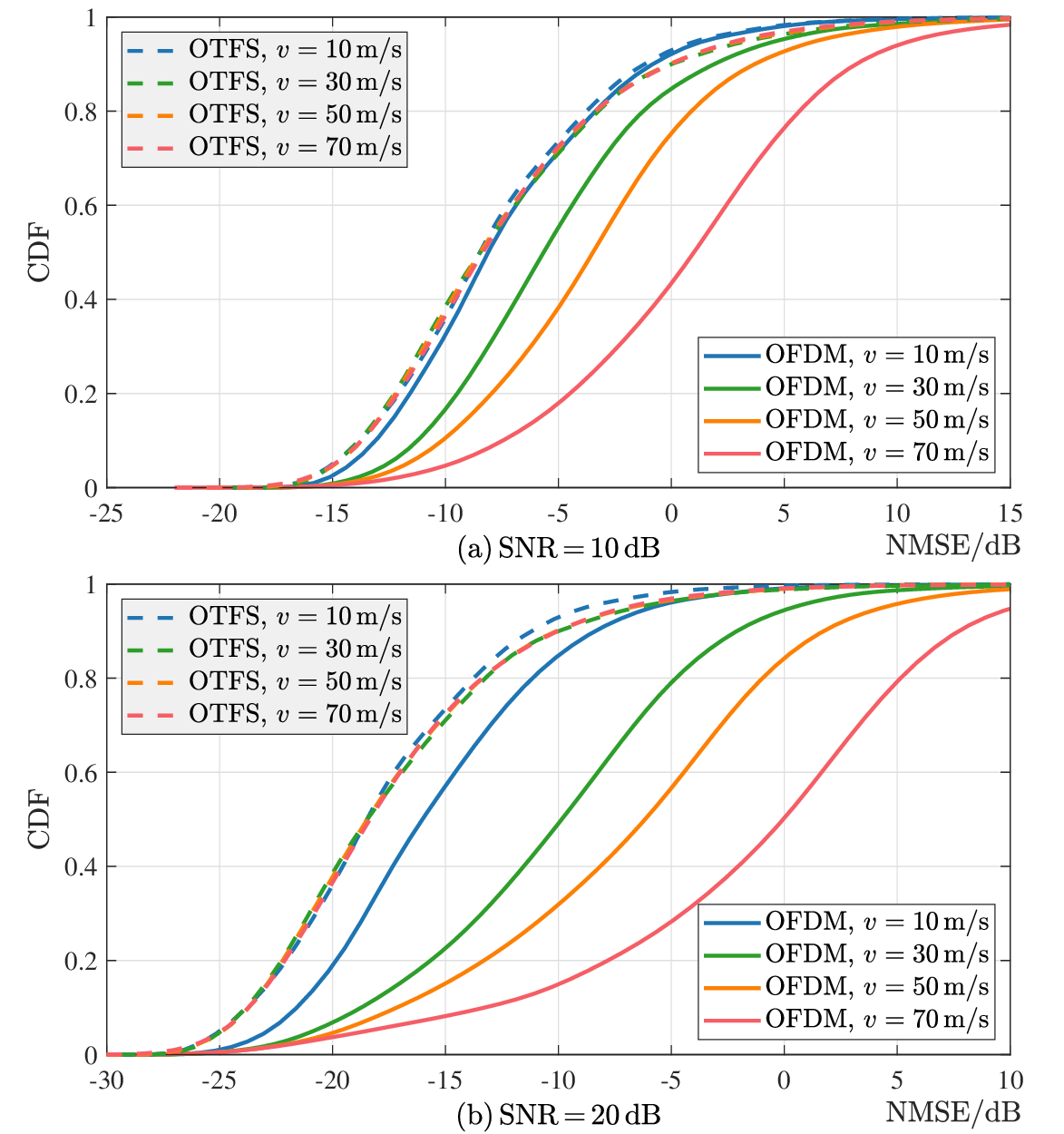}
	\caption{Normalized MSE of channel estimation errors for OTFS and OFDM with the same overhead.}
	\label{nmse}
\end{figure}

In Fig. \ref{nmse}, the normalized MSE of channel estimate is presented for both OFDM and OTFS, under different speeds. The x-axis is the achievable rate, while the y-axis is the cumulative density function (CDF). As we can see, the performance of OFDM is sensitive to the speed of the vehicle, while OTFS has similar performance in different speeds. 

The results demonstrate the robustness of OTFS to Doppler spread. As the speed increase, both OTFS and OFDM will see performance degradation. For OFDM, the performance degradation is severe for two reasons. First, the LTI channel model cannot describe the dynamics of the wireless channel, and the channel estimation error accumulates over time. Second, the dispersion in delay and Doppler lead to ISCI. For OTFS, we only see slight performance degradation, due to the increase ISCR from double dispersion.

\subsection{Spectral Efficiencies of OFDM and OTFS}
As we have mentioned at the beginning, the major problem of applying OFDM in mobile channels is the frequent channel estimation, which leads to significant overhead and reduced spectral efficiency. In this part, we will compare the ergodic achievable rates of OTFS and OFDM, by considering both the ISCI, channel training overhead and also the channel estimation error. The first step is to estimate the CSI, and the CSI will then be used for data detection. In this part, the bandwidth is chosen as $B=10$\,MHz. The delay spread is $\tau_D=300/c=1\upmu$s, and Doppler spread varies with the speed of the mobile device.

\begin{figure}[htp]
	\centering
	\includegraphics[width=0.45\textwidth]{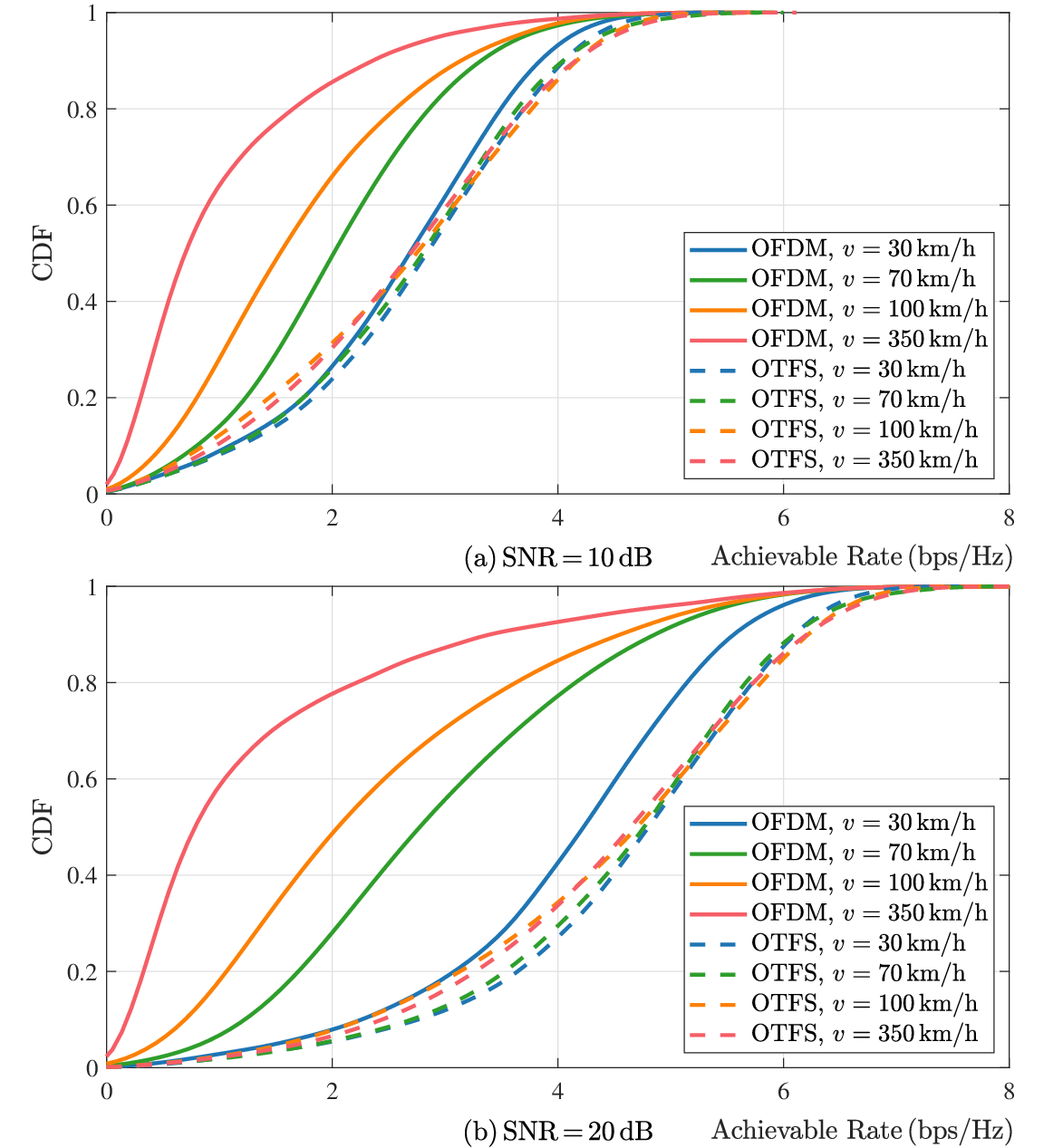}
	\caption{Achievable rates of OTFS and OFDM with multiple access. Both OTFS and OFDM are using the same amount of resources. The SER of OTFS will significantly outperform OFDM.}
	\label{achievableRate}
\end{figure}

The channel estimation error leads to reduced SINR, and thus reduced spectral efficiency. In Fig. \ref{achievableRate}, the achievable rates of OTFS and OFDM are presented. Specifically, the x-axis is the achievable rate, while the y-axis is the CDF.

Similar to Fig. \ref{nmse}, we can see that the OTFS has much better performance than OFDM. Besides, OFDM is very sensitive to channel mobility, while OFDM is much more robust. The fundamental reason is that the OTFS is based on the time-variant D-D domain channel model, which incorporated the channel dynamics in signal processing. In this case, we assume the OFDM and OTFS are using the same amount of resources for channel estimation. In this case, the OTFS can estimate the channel with much higher accuracy. What if we assign more resources for channel estimation in OFDM, so that the channel estimation accuracy is identical for both cases?

%
%

\section{Conclusions}\label{conclusions}

In this paper, we talked about the advantages of OTFS over OFDM in terms of spectral efficiency, resulting from the much reduced channel training overhead. We showed that the D-D domain channel model is also an approximation of the real channel, but it is more accurate then the LTI model, and thus allows us to estimate the channel with a much reduced channel estimation frequency. The predictability of the channel in T-F domain comes from the sparsity of response in D-D domain. Besides, we showed that it's possible to use a very small amount of resources for channel interpolation. A pipeline algorithm is proposed for channel interpolation with reduced processing delay. Further more, we showed that channel extrapolation and data-aided channel tracking would be possible, benefiting from the predictability of the D-D domain channel. Two sources of channel interpolation error are unveiled: the D-D domain aliasing resulting from the finite T-F window, and the ISCI induced by channel dispersion. Their impacts on channel estimation error are quantified. Overall, we can conclude that OTFS has a huge advantage over OFDM due to the reduced channel training overhead. As a matter of fact, this advantage actually comes from the D-D domain channel model itself, and is thus shared by other signaling techniques designed for doubly-dispersive channels.

\begin{appendices}
\section{Sum of $\sinc$ and Dirichlet Functions}
To start with, note that we can rewrite the periodical extension of $w(\tau,\nu)$ as the product of two sums of sinc functions:
\begin{equation}
\begin{split}
\ & \frac{1}{MN}\sum_{k,l} w(\tau-kT,\nu-lF)\\
=&\sum_{k} \sinc(\pi NT\nu -\pi lN)e^{-j\pi (N-1)T\nu}\cdot\\
\ & \sum_l\sinc(\pi M F(\tau-kT))e^{j\pi (M-1)F\tau}.
\end{split}
\end{equation}
The first sum can be rewritten as 
\begin{equation}
\begin{split}
\ &\sum_{l} \sinc(\pi NT\nu-\pi lN)e^{-j\pi (N-1)T(\nu-lF)}\\
=&e^{-j\pi (N-1)T\nu}\diric(N,2\pi T\nu),
	\end{split}
\end{equation}
and the equality is justified in \cite{Gong2024SignalInterp}.
Similarly, we have
\begin{equation}
\begin{split}
\ &\sum_{l} \sinc(\pi MF\tau-\pi kM)e^{j\pi (M-1)F(\tau-kT)}\\
=&e^{-j\pi (M-1)F\nu}\diric(M,2\pi F\tau).
	\end{split}
\end{equation}
This concludes the proof of \eqref{eqn016}.
%
\end{appendices}

\bibliographystyle{ieeetr}
\bibliography{DDMM_ChanEst}

\end{document}